\documentclass[aps,prl,twocolumn,superscriptaddress]{revtex4}

\usepackage[pdftex]{graphicx}% Include figure files
\usepackage{amsmath}
\usepackage{amssymb}

\begin{document}

\title{Multi-Orbital Quantum Phase Diffusion}

\author{Sebastian Will}
\affiliation{Institut f\"ur Physik, Johannes-Gutenberg Universit\"at, 55099 Mainz, Germany}
\affiliation{Ludwig-Maximilians-Universit\"at, 80799 M\"unchen, Germany}
\author{Thorsten Best}
\affiliation{Institut f\"ur Physik, Johannes-Gutenberg Universit\"at, 55099 Mainz, Germany}
\author{Ulrich Schneider}
\affiliation{Institut f\"ur Physik, Johannes-Gutenberg Universit\"at, 55099 Mainz, Germany}
\affiliation{Ludwig-Maximilians-Universit\"at, 80799 M\"unchen, Germany}
\author{Lucia Hackerm\"uller}
\affiliation{Institut f\"ur Physik, Johannes-Gutenberg Universit\"at, 55099 Mainz, Germany}
\author{Dirk-S\"oren L\"uhmann}
\affiliation{I.~Institut f\"ur Theoretische Physik, Universit\"at Hamburg, 20355 Hamburg, Germany}
\author{Immanuel Bloch}
\affiliation{Institut f\"ur Physik, Johannes-Gutenberg Universit\"at, 55099 Mainz, Germany}
\affiliation{Ludwig-Maximilians-Universit\"at, 80799 M\"unchen, Germany}
\affiliation{Max-Planck-Institut f\"ur Quantenoptik, 85748 Garching, Germany}

\date{\today}

\begin{abstract}
The collapses and revivals of a coherent matter wave field of interacting particles can serve as a sensitive interferometric probe of the interactions and the number statistics of the underlying quantum field. Here we show how the ability to observe long time traces of collapse and revival dynamics of a Bose-Einstein condensate loaded into a three-dimensional (3D) optical lattice allowed us to directly reveal the atom number statistics and the presence of effective coherent multi-particle interactions in a lattice. The multi-particle interactions are generated via virtual transitions to higher lattice orbitals and can find use for simulations of effective field theories with ultracold atoms in optical lattices \cite{Johnson:2009}. We measured their absolute strengths up to the case of six-particle interactions and compare our findings with theory.
\end{abstract}
\maketitle

%\cite{Johnson:2009}

Coherent quantum states represent the most robust and stable field solutions in physics \cite{Glauber:1963}. Characterized by a single amplitude and phase, they have found widespread use in the quantum description of classical coherent fields, ranging from laser light to coherent matter waves in superconductors, superfluids or atomic Bose-Einstein condensates. Whenever interactions between the underlying particles are present, or - more generally - whenever the phase evolution of the number states (Fock states) that form the coherent state is nonlinear in particle number, coherent states can undergo an intriguing sequence of collapses and revivals. In such a sequence, the quantum state first evolves into a highly correlated and entangled state where at the time of the collapse the classical field vanishes, however, at a later time the entanglement is unraveled again and the original classical field is ideally recreated.

%%%%%%%%%%%%%%%%%%%%%%%%%%%%%%%%%%%%%%%%%%%%%%%%%%%%%%%%%%%%%%%%%%%%%%%%%
% FIGURE 1
%%%%%%%%%%%%%%%%%%%%%%%%%%%%%%%%%%%%%%%%%%%%%%%%%%%%%%%%%%%%%%%%%%%%%%%%%

\begin{figure}[h!]
\begin{center}
\includegraphics[width=0.95\columnwidth]{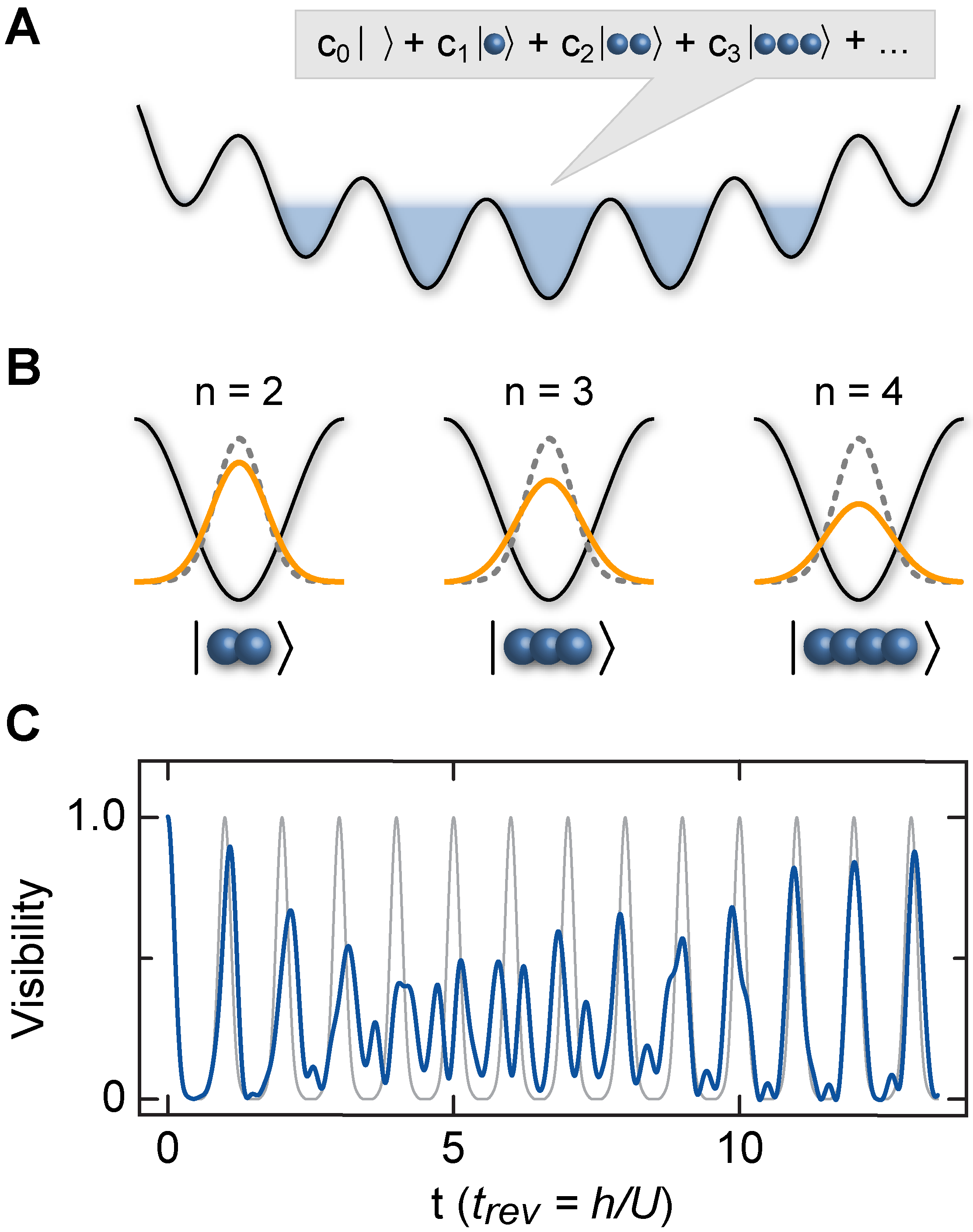}
\end{center}
\caption{\label{fig:cartoon}  (\textbf{A}) A Bose-Einstein condensate loaded into a weak optical lattice forms a superfluid with each atom being delocalized over several lattice sites. The quantum state on each site can be expressed as a superposition of Fock states $|n\rangle$ with amplitudes $c_n$. (\textbf{B}) For repulsive onsite interactions virtual transitions to higher lattice orbitals broaden the ground state wavefunction on a lattice site compared to the non-interacting system (gray dashed line). This gives rise to characteristic Fock state energies, which can be described by effective multi-particle interactions. (\textbf{C}) A coherent state of interacting atoms confined to a deep lattice well undergoes multi-orbital quantum phase diffusion (blue solid line). The beat signal indicates coherent multi-particle interactions. The dynamics are markedly different from the monochromatic evolution expected in a single-orbital model with a single two-particle interaction energy $U$ (gray solid line).}
\end{figure}

Remarkable examples of such collapses and revivals have been observed for a coherent light field interacting with a single atom in Cavity Quantum Electrodynamics \cite{Brune:1996,HarocheRaimondbook}, for a classical oscillation of a single ion held in a trap \cite{Meekhof:1996,Poschinger:2009} or for a matter wave field of a Bose-Einstein condensate \cite{Greiner:2002b,Anderlini:2006,Sebby:2007}. In the latter case, where the dynamics are induced by non-linear two-particle interactions between the atoms, the underlying process of quantum phase diffusion \cite{Yurke:1986,Wright:1996,Wright:1997,Walls:1997,Imamoglu:1997,Castin:1997} has been proposed to be useful for the creation of highly correlated and entangled quantum states, such as large Schr\"odinger cat or spin-squeezed states \cite{Sorensen:2001,Micheli:2003,Pezze:2009}, the latter of which have been demonstrated recently \cite{Esteve:2008}. In all these cases, it is typically assumed that the atoms occupy a single spatial orbital of the system. Atom-atom interactions can however promote particles to higher lying orbitals. The admixture of these excited orbitals modifies the shape of the spatial wavefunction and gives rise to renormalized interaction energies depending on the atom number, in close analogy to the so-called configuration interaction in quantum chemistry \cite{SzaboOstlundbook}. An instructive way to envisage the physical processes is provided by effective field theory: Virtual transitions of atoms to excited orbitals generate effective multi-particle interactions as higher order corrections to the single-orbital two-particle interaction \cite{Johnson:2009}. Previous experiments have only allowed the observation of few cycles of quantum phase diffusion dynamics \cite{Greiner:2002b,Anderlini:2006,Sebby:2007} and therefore could not reveal the striking signatures and consequences of multi-orbital effects, which are expected to lead to a multitude of novel many-body quantum phases \cite{Alon:2005, Zhao:2008}. Here, we have been able to observe more than 40 collapses and revivals in long time traces of the quantum evolution of coherent matter wave fields trapped in a 3D optical lattice potential. Multi-orbital effects give rise to a discrete set of distinct two-particle interaction energies for different Fock states constituting a coherent field. These in turn result in characteristic beat signals in the collapse and revival time traces, from which we infer the occupation of Fock states, measure their energies with high precision and determine the strength of the effective multi-particle interactions. Recently, multi-particle interactions have been identified via loss resonances for three- and four-body recombination \cite{Kraemer:2006,Knoop:2009,Zaccanti:2009}, however \emph{coherent} multi-particle interactions, which we demonstrate here in the form of effective interactions, have so far defied observation.

%\section{Theoretical Model}

Let us consider a single site of a deep optical lattice filled with $n$ atoms occupying the lowest single-particle state of the system $\psi_0(\mathbf{r})$. Assuming weak interactions and excluding multi-orbital effects, collisions between the atoms lead to an energy shift of the single-orbital Fock state energy given by $E_{n}^{SO}=U n(n-1) / 2$ representing the onsite interaction energy of the Bose-Hubbard model. Here $U=4\pi \hbar^2 a_{s} /m \int |\psi_0(\mathbf{r})|^4\, d^3r$ denotes the two-particle interaction energy, determined by the onsite wavefunction $\psi_0(\mathbf{r})$, the $s$-wave scattering length $a_{s}$ and the mass of an atom $m$. Within the restriction to the lowest vibrational state of the system, $U$ is independent of the filling $n$ at the lattice site.

Inducing virtual transitions to higher vibrational states, interactions modify the shape of the ground state wavefunction (Fig.~\ref{fig:cartoon}B) and the two-particle interaction energy $U$ itself becomes atom number dependent \cite{Luehmann:2008,Johnson:2009,Lutchyn:2009,Alon:2005,Li:2006,Hazzard:2009}. In this case, the multi-orbital Fock state energies in a single lattice well can be approximated by the expansion
\begin{equation}
 \begin{split}
E_{n}^{MO} & = \frac{1}{2}\,U_2 \, n (n-1) + \frac{1}{6}\, U_3 \, n (n-1)(n-2) \,+\\
&+ \frac{1}{24}\, U_4 \, n (n-1)(n-2)(n-3) + \ldots,
\label{eq:effint}
 \end{split}
\end{equation}
representing the eigenenergies of the corresponding effective Hamiltonian \cite{Johnson:2009}. Here, coherent multi-particle interactions become explicitly visible with the characteristic strength of the $m$-particle interactions being denoted by $U_{m}$.

%%%%%%%%%%%%%%%%%%%%%%%%%%%%%%%%%%%%%%%%%%%%%%%%%%%%%%%%%%%%%%%%%%%%%%%%%
% FIGURE 2
%%%%%%%%%%%%%%%%%%%%%%%%%%%%%%%%%%%%%%%%%%%%%%%%%%%%%%%%%%%%%%%%%%%%%%%%%

\begin{figure*}
\begin{center}
\includegraphics[width=1.33\columnwidth]{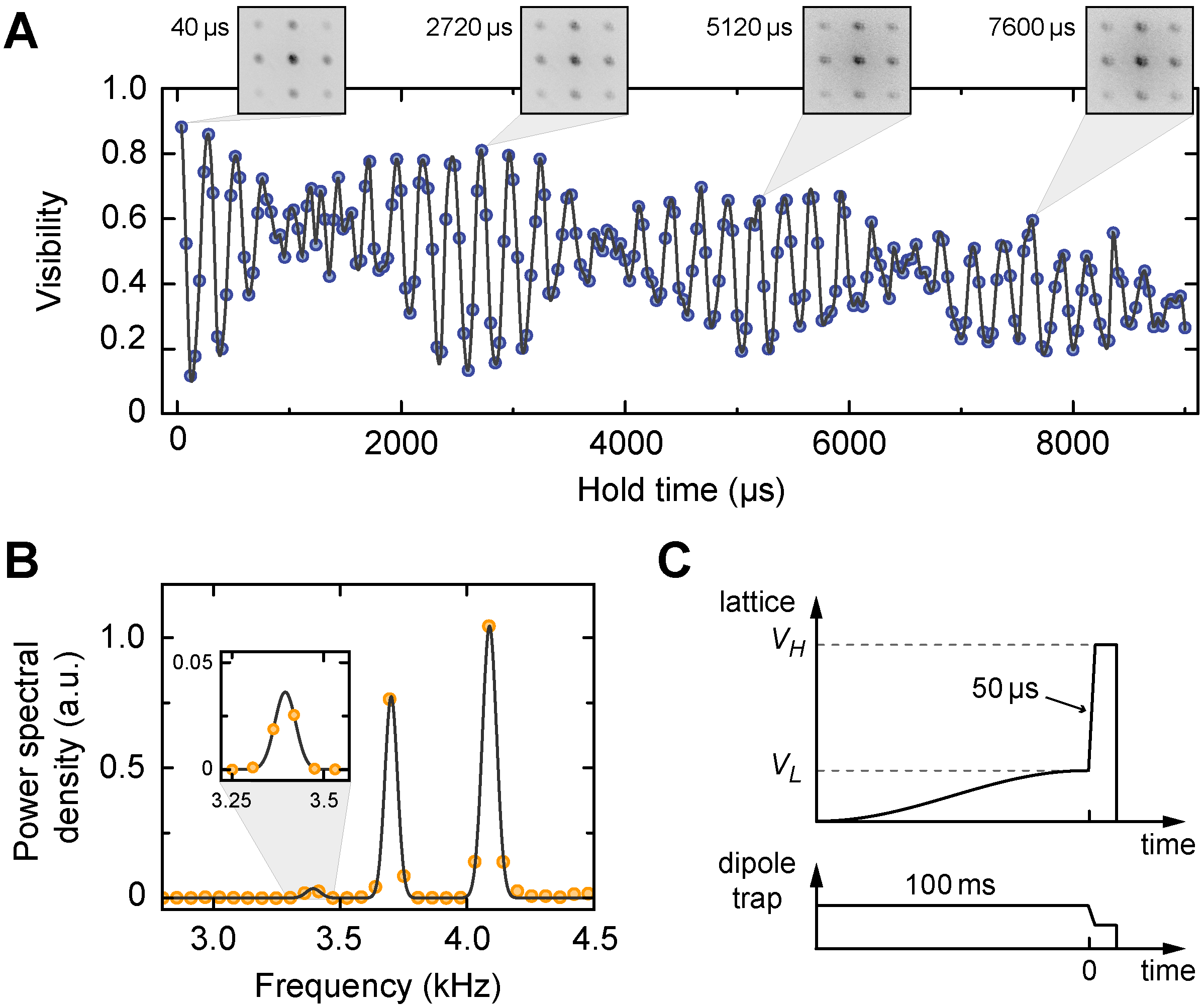}
\end{center}
\caption{\label{fig:longtrace} \textbf{Multi-orbital quantum phase diffusion of an atom number superposition state.} (\textbf{A}) The collapse and revival dynamics of a number-squeezed superposition state in a deep optical lattice is shown. A BEC of about $1.9 \pm 0.3 \times 10^5$ Rb atoms was adiabatically loaded to a $V_{L}=8$ $E_{rec}$ lattice within $100$ ms. Phase evolution was induced by a non-adiabatic jump into a $V_{H}=41$ $E_{rec}$ deep lattice, preserving superposition states with finite number fluctuations and an ensemble averaged (central) mean atom number of about $\langle \bar{n} \rangle = 1.0$ ($\bar{n}_{c} = 2.5$). Simultaneously with the lattice jump, the underlying harmonic confinement was instantaneously minimized (\textbf{C}) to optimize the coherence time. The phase evolution shows a beat note signature resulting from coherent multi-particle interactions with different interaction strengths. Each data point corresponds to a single run of the experiment. The solid line interpolates the data and serves as a guide to the eye. (\textbf{B}) Spectral analysis of the time trace (A) using a numerical Fourier transform reveals the contributing frequencies. The solid line shows Gaussian fits to the peaks.}
\end{figure*}

%\section{Theory of quantum phase diffusion}

An efficient way to experimentally probe the eigenenergies of a Hamiltonian, is to monitor the non-equilibrium dynamics of a quantum state prepared in a superposition of different eigenstates. In our case, the state is a superposition of different atomic Fock states  $|n\rangle$ and can be expressed in the general form $|\psi(t)\rangle = \sum _{n=0}^{\infty} c_{n} e^{-i E_{n} t / \hbar} |n\rangle$. Experimentally, a 3D array of such states is created by loading a BEC into a shallow 3D lattice potential (Fig.~\ref{fig:cartoon}A). The time evolution of the ensemble can be probed by analyzing the visibility of the atomic interference pattern as observed after rapid switch-off of the lattice potential and subsequent time-of-flight expansion \cite{Gerbier:2005a}. For an array of identical states $|\psi (t) \rangle$, the visibility of the interference pattern relates monotonically to $|\langle \psi (t)|\hat{a}|\psi(t)\rangle|^2$, where $\hat{a}$ denotes the annihilation operator at a lattice site. Hence, the dynamical evolution of the matter wave field, the quantum phase diffusion of $|\psi(t)\rangle$, is given by:
\begin{equation}
 \begin{split}
\label{eq:phdiff}
|\langle \psi & (t)|\hat{a}|\psi(t)\rangle|^2 \equiv |\langle \hat{a} \rangle|^2  = \sum_{m,n=0}^{\infty} \sqrt{m+1}\,\sqrt{n+1}\,\,\times  \\
&\times c_{m} \,c_{m+1}^{\ast}\, c_{n}^{\ast} \, c_{n+1} \, e^{i (E_{m+1} - E_{m}-E_{n+1} + E_{n}) t/\hbar}.
 \end{split}
\end{equation}
To illustrate the essential dynamics of Eq.~(\ref{eq:phdiff}), we assume the initial state $|\psi\rangle$ to be a coherent state with $c_{n}=e^{-|\alpha|^2/2}\, \frac{\alpha^n}{\sqrt{n!}}$, where $\alpha =\sqrt{\bar{n}}\cdot e^{i\phi}$ denotes the complex field amplitude with the mean atom number $\bar{n}$ and initial phase $\phi$. In a single-orbital model with eigenenergies $E_{n}^{SO}=U n(n-1)/2$, periodic collapses and revivals of the visibility are expected, which can analytically be expressed by $|\langle \hat{a} \rangle|^2/\bar{n} = e^{2\,\bar{n} (\cos( U t /\hbar)-1)}$ featuring a single frequency $U/h$ and its higher harmonics (Fig.~\ref{fig:cartoon}C (gray solid line)) \cite{Imamoglu:1997,Castin:1997}.

Multi-orbital effects, however, reach beyond the picture of monochromatic matter wave dynamics: the time-evolution of $|\langle\hat{a}\rangle|^2$ contains multiple frequency components since Fock state energies are no longer integer multiples of $U$, as shown in Fig.~\ref{fig:cartoon}C (blue solid line). Assuming repulsive interactions, we expect frequencies of order $\mathcal{O}(U/h)$ to be given by $E_2 / h > (E_3 - 2 E_2)/h > (E_4 -2 E_3 + E_2)/h > (E_5 - 2 E_4 + E_3)/h > \dots$ as well as higher order frequency components $\mathcal{O}(\ell \cdot U/h)$, with $\ell$ being a positive integer.

The detection of multi-orbital quantum phase diffusion over long times results in a high spectral resolution, allowing for a precise measurement of individual Fock state energies $E_n^{MO}$ and multi-particle interaction strengths. In addition, Eq.~(\ref{eq:phdiff}) implies that the spectral weight of each contributing frequency is directly connected to the coefficients $c_n$ of the initial state $|\psi \rangle$. Hence, the statistical contribution of single Fock states to the global superfluid quantum many-body state can be obtained complementary to previous work \cite{Gerbier:2006a, Campbell:2006,Cheinet:2008}.

%%%%%%%%%%%%%%%%%%%%%%%%%%%%%%%%%%%%%%%%%%%%%%%%%%%%%%%%%%%%%%%%%%%%%%%%%
% FIGURE 3
%%%%%%%%%%%%%%%%%%%%%%%%%%%%%%%%%%%%%%%%%%%%%%%%%%%%%%%%%%%%%%%%%%%%%%%%%

\begin{figure*}
\begin{center}
\includegraphics[width=1.9\columnwidth]{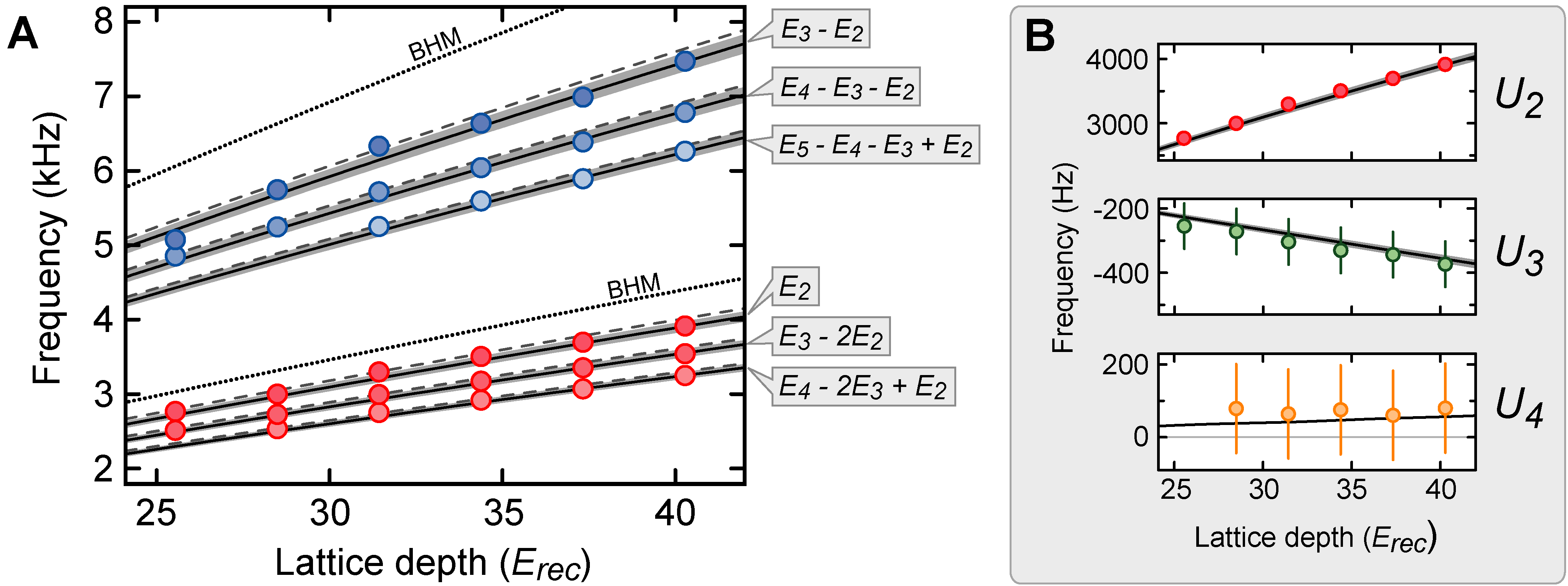}
\end{center}
\caption{\label{fig:absoluteubb} \textbf{Multi-orbital energies and multi-particle interactions.}
(\textbf{A}) Long collapse and revival traces were recorded under identical loading conditions ($V_L=8$ $E_{rec}$), but variable lattice depths $V_H$ during phase evolution. Numerical Fourier transform of the time traces reveals the contributing frequencies of orders $\mathcal{O}(U/h)$ (red circles) and $\mathcal{O}(2 U/h)$ (blue circles) with a typical uncertainty of $\pm 50$ Hz, the shading of the data points reflects the relative spectral weight. The solid lines with gray shading denote the theoretically expected frequencies at an $s$-wave scattering length $a_s = 102 \pm 2$ $a_0$ as derived for a basis set with $4^3$ orbitals. Calculation on a smaller basis set with $3^3$ orbitals yields slightly higher energies (dashed lines). The black dotted lines show the single-orbital interaction energy of the Bose-Hubbard model (BHM). At low lattice depths, only the strongest peaks can be resolved mainly due to smaller peak spacings. (\textbf{B}) Effective two-, three- and four-particle interaction strengths as derived from experiment and theory ($a_s = 102 \pm 2$ $a_0$, $4^3$ orbitals). The error bars correspond to standard propagation of uncertainty. }
\end{figure*}

%\section{Experimental Setup and Sequence}

Our experiments began with an atomic Bose-Einstein condensate of repulsively interacting $^{87}$Rb atoms in the $|F=1,m_F=+1\rangle$ hyperfine state, with variable atom numbers between $1.2 \times 10^5$ and $3.5 \times 10^5$. The atoms were initially held in a pancake-shaped crossed optical dipole trap at a wavelength of $\lambda_{dip}=1030$\,nm with trap frequencies $\omega_z= 2 \pi \times 130$ Hz in the direction of gravity and $\omega_{\perp}= 2 \pi \times 32$ Hz in the orthogonal plane. Subsequently a 3D optical lattice ($\lambda=738$\,nm) of simple cubic type was adiabatically ramped up to lattice depths $V_L$ between $3-13\,E_{rec}$, where $E_{rec}=h^2/(2 m \lambda^2)$ denotes the recoil energy. For these lattice depths the many-body ground state of the system is a superfluid and we expect the states on a lattice site to range from coherent states for shallow lattice depths to highly number squeezed states for deeper lattices \cite{Greiner:2002b,Gerbier:2006a,Esteve:2008,Cheinet:2008}.

A sudden increase of the lattice depth from $V_L$ to $V_H$ ranging between $25-41\,E_{rec}$ strongly suppressed the tunnel coupling and essentially froze out the atom number distribution on each site. In this regime, the time evolution of each site is governed by the Fock state energies of Eq.~(\ref{eq:effint}) and the quantum phase diffusion process started. The jump to $V_H$ was carried out within $50$ $\mathrm{\mu s}$ which is slow enough to avoid populating higher vibrational states, but fast compared to tunneling within the first band. Previous experiments were limited to follow quantum phase diffusion only for short traces with a maximum of 4-5 collapse and revival cycles \cite{Greiner:2002b,Anderlini:2006,Sebby:2007} since a global harmonic confinement led to rapid relative dephasing of lattice sites. Here, the different detunings of the optical lattice (blue-detuned) and the dipole trap (red-detuned) with respect to the atomic resonance ($\lambda_{res}=780$ nm) allowed us to change the underlying harmonic confinement during the experimental sequence (Fig.~\ref{fig:longtrace}C). Along with the jump to $V_H$ we instantaneously reduced the dipole trap in order to cancel harmonic confinement in the horizontal plane, boosting the coherence time of quantum phase diffusion. Under such conditions, the atomic sample stayed trapped due to the presence of the deep optical lattice $V_H$. After letting the system evolve for a hold time $t$, all trapping potentials were switched-off simultaneously and an absorption image of the matter wave interference pattern was recorded after 10 ms time-of-flight. The visibility of the interference pattern was evaluated, being a measure for the ensemble average of $|\langle \hat{a} \rangle|^2$ \cite{Gerbier:2005a}.

%\section{General signal}

A typical time trace ($V_L=8\,E_{rec}$) of quantum phase diffusion is shown in Fig.~\ref{fig:longtrace}A displaying about 40 revivals. On top of a fast series of collapses and revivals, we observe a slower modulation of the envelope indicating a beat between at least two different energies in the system. The spectral content of the time trace has been quantitatively analyzed using a numerical Fourier transform of the data (Fig.~\ref{fig:longtrace}B). We clearly identify three distinct frequency components of order $\mathcal{O}(U/h)$ present in the time trace, the smallest one originating from sites occupied by up to four atoms. It is remarkable that our technique can reveal even very small Fock state amplitudes $c_n$ due to a heterodyne effect with other Fock states $|n-1\rangle$ and $|n-2\rangle$ of typically larger amplitudes $c_{n-1}$ and $c_{n-2}$ (Eq.~\ref{eq:phdiff}). The residual damping on the time trace most likely results from residual harmonic confinement (along the $z$-direction), residual tunneling \cite{Fischer:2008} via higher bands and the finite extension of the atomic ensemble, sampling a slightly inhomogeneous distribution of lattice depths due to the Gaussian shape of the lattice laser beams.

%%%%%%%%%%%%%%%%%%%%%%%%%%%%%%%%%%%%%%%%%%%%%%%%%%%%%%%%%%%%%%%%%%%%%%%%%
% FIGURE 4
%%%%%%%%%%%%%%%%%%%%%%%%%%%%%%%%%%%%%%%%%%%%%%%%%%%%%%%%%%%%%%%%%%%%%%%%%

%\begin{widetext}
\begin{figure*}
\begin{center}
\includegraphics[width=1.64\columnwidth]{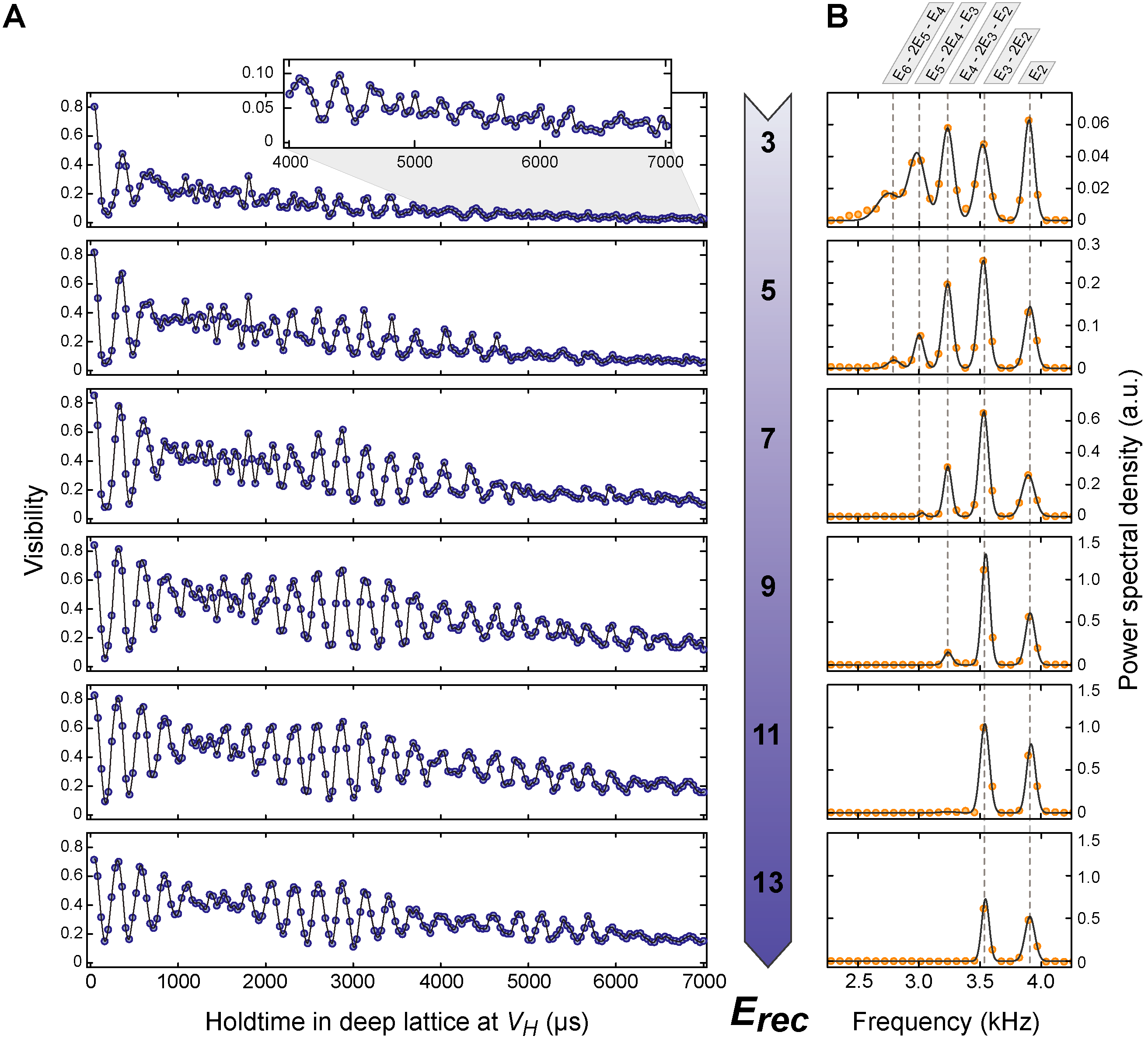}
\end{center}
\caption{\label{fig:squeezingdata} \textbf{Global number statistics upon approaching the Mott insulator transition.}
 (\textbf{A}) Multi-orbital quantum phase diffusion in a $V_H = 40$ $E_{rec}$ deep lattice was used after $3.3 \pm 0.3 \times 10^5$ ${}^{87}$Rb atoms had been adiabatically loaded to lattices ranging from $V_L=3$ to $13$ $E_{rec}$. The mean atom number of the individual traces differed by as little as $\pm 1$\%. Although the coherence time at shallow lattices is significantly reduced, the visibility reliably shows dynamics down to the percent level (inset). (\textbf{B}) The corresponding Fourier power spectra reveal frequency contributions with up to six-particle Fock state energies. The peak positions agree with the theoretical prediction (dashed vertical lines) and are independent of $V_L$. Number-squeezing manifests itself in both reduced peak amplitudes and a narrowing of the spectra for increasing $V_L$. The solid lines show Gaussian fits to the peaks.}
\end{figure*}
%\end{widetext}

%\section{Detection of renormalized interaction energies}

In order to map out the dependence of the multi-orbital Fock state energies $E_n^{MO}$ on lattice depth, we recorded several long collapse and revival traces with $V_H$ between $25 - 41$~$E_{rec}$ (Fig.~\ref{fig:absoluteubb}). With a relative statistical uncertainty on the level of few percent, a Fourier analysis reveals both a set of frequencies of order $\mathcal{O}(U/h)$ and $\mathcal{O}(2 U/h)$, monotonically increasing with lattice depth. The measured two-particle Fock state energy $E_2$ is about $10 \pm 1$ \% lower than predicted by a simple single-orbital Bose-Hubbard model, which strongly indicates the relevance of multi-orbital effects in optical lattices (Fig.~\ref{fig:absoluteubb}A).

We compare our experimental data to an \emph{ab-initio} exact diagonalization of a single-site multi-orbital system. For this system a contact interaction with an $s$-wave scattering length of $102 \pm 2$ $a_0$ \cite{Klausen:2001} was assumed, where $a_0$ denotes the Bohr radius. We find excellent agreement, particularly for deep lattices ($V_H > 34$ $E_{rec}$), when the diagonalization is performed on a Hilbert-space with $4^3$ single-particle orbitals, corresponding to the four lowest lattice bands in three dimensions (Fig.~\ref{fig:absoluteubb}A, black solid lines). For lower lattice depths ($V_H < 34$ $E_{rec}$) the results of a calculation including $3^3$ onsite orbitals appear even more suitable (Fig.~\ref{fig:absoluteubb}A, gray dashed lines). The remarkable congruence between theory and experiment suggests, that the influence and occupation of even higher lying orbitals is negligible.

The strength of the effective multi-particle interactions has been iteratively derived from the measured multi-orbital energies $E_n^{MO}$ according to $U_2=E_2^{MO}$, $U_3=E_3^{MO}-3 U_2$ and $U_4=E_4^{MO}-6 U_2 -4 U_3$ (Fig.~\ref{fig:absoluteubb}B). The resulting effective three-particle interaction is attractive and the measured values agree well with the results obtained from exact diagonalization, indicating that direct three-particle interactions are negligible for our experimental parameters. The measured effective four-particle interaction strengths on a scale as low as $100$ Hz are also in good agreement with theory, taking the experimental uncertainty into account. Our data shows, that the expansion provided by Eq.~\ref{eq:effint} quickly converges, offering the possibility to efficiently incorporate the effects of multi-orbital physics in refined effective single-band lattice Hamiltonians.

%\section{Measuring the atomic distribution}

In addition to its capability to measure coherent interatomic interaction strengths, multi-orbital quantum phase diffusion can be used to reveal the number statistics of many-body quantum states:  The spectral weight of the detected frequencies carries information on the Fock state amplitudes $c_n$ (Eq.~\ref{eq:phdiff}). To demonstrate this, we adiabatically prepared 3D arrays of coherent ($V_L = 3$ $E_{rec}$) to highly number-squeezed states ($V_L = 13$ $E_{rec}$) close to the Mott-insulator transition around $15$ $E_{rec}$, and employed quantum phase diffusion as a detection sequence (see Fig.~\ref{fig:squeezingdata}A). From $3$ to $11\,E_{rec}$ the time traces evolve from seemingly irregular oscillations into a clear beat signal since less frequencies contribute, as can be observed in the corresponding Fourier spectra (Fig.~\ref{fig:squeezingdata}B). This narrowing of the spectrum reflects a decreasing variance of the atom number distribution caused both by number squeezing and the reduction of the average onsite density due to increased interactions when the lattice depth is raised.

%\section{Summary and Outlook}

In conclusion, we have demonstrated how interaction induced virtual transitions to higher lattice orbitals lead to effective coherent multi-particle  interactions \cite{Johnson:2009}. We have determined their strengths by monitoring long time traces of the collapse and revival dynamics of the matter wave field of a BEC. Our measurements explicitly demonstrate the coherence of the multi-particle interactions and the number state resolving capabilities of multi-orbital quantum phase diffusion. Multi-orbital quantum phase diffusion can serve as a precise experimental technique to determine renormalized interaction energies in lattice based quantum gases, which have been found to play a crucial role in more complex systems such as Bose-Fermi mixtures \cite{Luehmann:2008,Best:2009,Lutchyn:2009}. Our results also show that multi-orbital effects will play an important role in experiments working towards interaction induced dynamical creation of spin-squeezing or Schr\"odinger cat states with ultracold atoms. Furthermore, we envisage that the coherent multi-particle interactions demonstrated here could be employed to simulate effective field theories \cite{Johnson:2009} and also help in realizing novel strongly correlated many-body quantum phases \cite{Alon:2005,Hazzard:2009}, e.g.~with topological order \cite{Paredes:2007} or exotic ground state properties \cite{Capogrosso:2009}.

This work was supported by the DFG, the EU (IP SCALA), EuroQUAM (LH), DARPA (OLE program), AFOSR, MATCOR (SW) and the Gutenberg Akademie (SW)


% If in two-column mode, this environment will change to single-column
% format so that long equations can be displayed. Use
%\begin{widetext}
%\end{widetext}

% figures should be put into the text as floats.
% Use the graphics or graphicx packages (distributed with LaTeX2e)
% and the \includegraphics macro defined in those packages.
% See the LaTeX Graphics Companion by Michel Goosens, Sebastian Rahtz,
% entire page. There is no need to do explicit centering.

% \begin{figure}
% \end{figure}

% Surround figure environment with turnpage environment for landscape
% figure
% \begin{figure}
% \end{figure}

%\appendix
%\section{}

%\begin{acknowledgments}
%\end{acknowledgments}


The authors declare that they have no competing financial interests.

Correspondence and requests for materials should be addressed to S.~Will (email: sebastian.will@lmu.de)

\end{document}